\documentclass[prd,preprint,superscriptaddress,amsmath,amssymb]{revtex4}

 \pdfoutput=1

\usepackage{graphicx}
\usepackage{color,slashed}

\newcommand{\oovt}{\frac{1}{3}}

\begin{document}
\title{\bf Flavor anomalies in leptoquark model with gauged  $U(1)_{L_\mu-L_\tau}$}

\author{Chuan-Hung Chen}
\email[E-mail: ]{physchen@mail.ncku.edu.tw}
\affiliation{Department of Physics, National Cheng-Kung University, Tainan 70101, Taiwan}
\affiliation{Physics Division, National Center for Theoretical Sciences, Taipei 10617, Taiwan}

\author{Cheng-Wei Chiang}
\email[E-mail: ]{chengwei@phys.ntu.edu.tw}
\affiliation{Department of Physics and Center for Theoretical Physics, National Taiwan University, Taipei 10617, Taiwan}
\affiliation{Physics Division, National Center for Theoretical Sciences, Taipei 10617, Taiwan}

\date{\today}

\begin{abstract}
Leptoquarks (LQs) have been extensively studied in the context of $B$ anomalies. When $U(1)_{L_\mu-L_\tau}$ is introduced to a scalar LQ model with the LQ $S_1$ charged under the new symmetry, $S_1$ primarily couples to the third-generation leptons while its couplings to first and second-generation leptons are naturally suppressed.  Furthermore, only $S_1$ in the scalar LQ models has the feature that down-type quarks merely couple to neutrinos but not the charged leptons, avoiding strict restrictions from $b\to s \mu^+ \mu^-$.  With this distinctive characteristic of $S_1$, we investigate its impact on rare processes involving the $d_i \to d_j \nu \bar\nu$ transitions.  Under the dominant constraints from $\Delta F=2$ processes, we find that the $S_1$ contributions to the branching ratios (BRs) of $B\to K(K^*) \nu \bar\nu$ and $K_L \to \pi^0 \nu \bar\nu$ can be factorized into the same multiplicative factor multiplying the standard model predictions.  Enhancement in the BRs can possibly exceed a factor of 2.  In particular, ${\cal B}(K^+\to \pi^+ \nu \bar\nu)$ can reach the upper $1\sigma$ error of the experimental value, i.e., $\simeq 15.4 \times 10^{-11}$.  We also show that the model can fit the new world averages of $R(D)$ and $R(D^*)$. 

\end{abstract}
\maketitle

\section{Introduction}

Loop-induced rare processes in the standard model (SM) are commonly considered promising places for probing new physics effects. One example is the muon anomalous magnetic dipole moment (muon $g-2$), which shows a $5.1\sigma$ deviation from the SM prediction~\cite{Muong-2:2023cdq, Aoyama:2020ynm}.  Using the exclusive- and hadronic-tag approaches with 362 fb$^{-1}$ of data, the Belle II Collaboration has observed the first evidence of $B^+\to K^+ \nu\bar\nu$ decay, which arises from the electroweak box and penguin diagrams in the SM. The combined result from both tag approaches is reported as~\cite{EPS_Belle2a}:
 \begin{align}
 {\cal B}(B^+ \to K^+ \nu \bar\nu) =( 2.4 \pm 0.7)\times 10^{-5}\,,
 \end{align}
indicating a $2.8\sigma$ deviation from the SM prediction.
When combined with earlier measurements by BaBar~\cite{BaBar:2013npw} and Belle~\cite{Belle:2013tnz,Belle:2017oht,Belle-II:2021rof}, the observed branching ratio (BR) becomes $ {\cal B}(B^+ \to K^+ \nu \bar\nu) =(1.4\pm 0.4)\times 10^{-5}$. If we take the SM prediction to be ${\cal B}(B^+\to K^+ \nu \bar\nu)^{\rm SM}=(4.42 \pm 0.60) \times 10^{-6}$~\cite{Buras:2022wpw}, the ratio of the measurement to the SM result can be estimated as:
 \begin{equation}
 R^\nu \equiv \frac{{\cal B}(B^+\to K^+ \nu\bar\nu)}{{\cal B}(B^+\to K^+ \nu\bar\nu)^{\rm SM}}=3.17 \pm 1.00\,.
 \label{eq:BelleII-data}
 \end{equation}
This deviation from the SM prediction hints at the possibility of exotic interactions in $b\to s \nu \bar\nu$ or $b\to s + \text{invisible}$ processes~\cite{Buras:2014fpa,Browder:2021hbl,Asadi:2023ucx,Athron:2023hmz,Bause:2023mfe,Allwicher:2023syp,Felkl:2023ayn,Dreiner:2023cms,Amhis:2023mpj}. In this study, we focus on the former scenario and, more generally, we investigate the rare decaying processes involving the $d_i \to d_j \nu \bar\nu$ transitions, where $(d_{i}, d_j) = (b, s)$ or $(s, d)$.

Leptoquarks (LQs) have been broadly studied as potential solutions to the anomalies of lepton-flavor universality measured in $B$ meson decays~\cite{Fajfer:2012jt,Sakaki:2013bfa,Calibbi:2015kma,Sahoo:2015qha,Chen:2017hir,Crivellin:2019dwb,Davighi:2020qqa,Greljo:2021xmg,Carvunis:2021dss,Davighi:2022qgb,Heeck:2022znj}. Among the scalar LQ models, the LQ $S_1$ with the $SU(3)_C \times SU(2)_L \times U(1)_Y$ quantum numbers $(3,1,-2/3)$ couples down-type quarks to neutrinos but not charged leptons. Due to this distinctive feature, the effects of $S_1$ only affect the $d_i\to d_j \nu \bar\nu$ processes but not those involving the $d_i \to d_j \ell^+ \ell^{\prime-}$ transitions.  The fact that current experimental measurements on $B_s \to \mu^+ \mu^-$~\cite{ParticleDataGroup:2022pth,HeavyFlavorAveragingGroup:2022wzx} and $R(K^{(*)})$~\cite{LHCb:2022vje} show no significant deviations from the SM~\cite{Buras:2022qip} makes the $S_1$ model a perfect model to explain the above mentioned $ R^\nu$ anomaly and to enhance the BRs of $B\to K(K^*) \nu \bar\nu$ in general.

In addition to the BR enhancements in the $B\to K (K^*) \nu \bar\nu$ decays, the $S_1$ model can also be used to resolve the $R(D^{(*)})$ anomalies~\cite{Chen:2017hir}. The current experimental values are $R(D)=0.357\pm 0.029$ and $R(D^*)=0.284\pm 0.012$~\cite{HeavyFlavorAveragingGroup:2022wzx}, while the SM predictions are $R(D)\approx 0.30$ and $R(D^*)\approx 0.25$~\cite{MILC:2015uhg,Na:2015kha,Bigi:2016mdz,Bernlochner:2017jka,Jaiswal:2017rve,BaBar:2019vpl,Bordone:2019vic,Martinelli:2021onb}. These measurements indicate a notable $3.3\sigma$ deviation from the SM in the $b\to c \tau \nu$ decays~\cite{HeavyFlavorAveragingGroup:2022wzx}.

Without imposing further symmetry, the $S_1$ couplings to quarks and neutrinos are generally quark-flavor and lepton-flavor dependent. It is thus a common practice in the literature that an arbitrary structure of the flavor couplings is assumed.  Here we propose that a structure can be naturally obtained by imposing a gauge symmetry.  Models with the local $U(1)_{L_\mu -L_\tau}$ gauge symmetry, denoted by $U(1)_{\mu-\tau}$, have been extensively studied for various phenomenological reasons~\cite{He:1991qd,Heeck:2011wj,Chen:2017cic}, including its potential role in resolving the muon $g-2$ anomaly~\cite{Altmannshofer:2014cfa,Altmannshofer:2014pba,Altmannshofer:2016oaq,Chen:2023mep}. Once the SM symmetry is extended to include the $U(1)_{\mu-\tau}$ gauge symmetry, the model then has the following features: (i) $S_1$ primarily couples to the third-generation leptons, while its couplings to first and second-generation leptons are naturally suppressed. (ii) In the absence of a new weak CP phase, there are only three independent down-type quark couplings in the model, denoted by $y^q_{Lk}$, which are interconnected by the Cabibbo-Kobayashi-Maskawa (CKM) matrix. This results in various flavor-changing neutral current (FCNC) processes in $B$ and $K$ decays involving these three parameters. (iii) The charged lepton mass matrix is forced to be diagonal due to the presence of the $U(1)_{\mu-\tau}$ gauge symmetry. Therefore, no lepton flavor mixings are introduced if the SM Higgs doublet is the only scalar field responsible for the spontaneous electroweak symmetry breakdown.

Since the effective Hamiltonian for the $d_i\to d_j \nu \bar\nu$ transitions mediated by $S_1$ has the same interaction structure as in the SM, the BRs of $B\to K(K^*) \nu \bar\nu$ and $K_L\to \pi \nu \bar\nu$ in this model can be factorized into a scalar factor, which encodes the effects of $S_1$, multiplied by the SM values. When considering the stringent constraints from the $\Delta F=2$ processes ($F=K$ or $B_q$), the typical values of $y_{Lk}^q$ are $y^q_{L1}\sim \lambda^2$ and $y^{q}_{L2}\sim \lambda/2$ when we set $y^q_{L3}\sim O(1)$. With this structure of the new Yukawa couplings, the BRs for $B\to K(K^*) \nu \bar\nu$ and $K_L\to \pi \nu \bar\nu$ can possibly exceed the SM predictions by at least a factor of 2. In this case, ${\cal B}(K^+\to \pi^+ \nu \bar\nu)$ can reach the upper $1\sigma$ error of the experimental value. In addition, $R(D)$ and $R(D^{*})$ can be enhanced up to the central values of current data.

In the following, we will formulate the BRs for the exclusive $d_i \to d_j \nu \bar\nu$ processes mediated by $S_1$ in Sec.~\ref{sec:B&K}. Constraints from $\Delta K=2$ and $\Delta B_q=2$ are analyzed in Sec.~\ref{sec:DF=2}. Detailed numerical analysis and discussions are given in Sec.~\ref{sec:NA}. Our findings are summarized in Sec.~\ref{sec:summary}.

\section{$B\to K^{(*)} \nu \bar\nu$ and $K\to \pi \nu \bar\nu$ via LQ $S_1$} \label{sec:B&K}

Under the assumed $U(1)_{\mu-\tau}$ gauge transformations, only the second- and third-generation leptons and the $S_1$ field transform in the following way:
 \begin{equation}
 (L_{\mu(\tau)}, \mu_R(\tau_R))  \to e^{\pm  i \theta_X } (L_{\mu(\tau)}, \mu_R(\tau_R)) \,, ~S^{-1/3}_1  \to  e^{i \theta_X } S^{-1/3}_1 \,.
 \end{equation}
The Yukawa interaction terms of the LQ $S_1$ are given by:
\begin{align}
-{\cal L}_{Y} & \supset  \overline{Q^c_L} i \tau_2  {\bf y}^q_L L_\tau (S^{-\oovt}_1)^* + \overline{u^c_R} {\bf y}^u_R \tau_R (S^{-\oovt}_1 )^*+ \text{H.c.}
~,
\label{eq:LY}
\end{align}
where the quark-flavor indices are suppressed, $Q^T_L=(u, d)_L$ and $L^T_\tau=(\nu_\tau ,\tau)_L$ represent the quark and the third-generation lepton doublets, respectively, and $F^c=C\gamma^0 F^*$  with $C$ being the charge conjugation operator. The $U(1)_{\mu-\tau}$ gauge symmetry restricts the charged lepton mass matrix to be diagonal; therefore, no lepton flavor mixings are introduced. Clearly, the LQ only couples to the third-generation leptons. Due to the lack of evidence that calls for new CP-violating sources in the processes considered in this work, CP violation originates purely from the Kobayashi-Maskawa (KM) phase in this study.  Therefore, ${\bf y}^q_L$ and ${\bf y}^u_R$ are assumed to be real parameters. Taking the up-type quarks to be the diagonalized states,  Eq.~(\ref{eq:LY}) in terms of physical states can be expressed as:
 \begin{align}
- {\cal L}_Y \supset   \left( \overline{u^C_L} {\bf y}^q_L  P_L \tau+ \overline{u^C_R} {\bf y}^u_R P_R  \tau \right) (S^{-\oovt})^* - \overline{d^C_L} V^T {\bf y}^q_L P_L \nu_{\tau} (S^{-\oovt} )^*+ \text {H.c.}\,, \label{eq:LQ_C}
 \end{align}
with $V= V^{d\dag}_L$ being the CKM matrix.

From Eq.~(\ref{eq:LQ_C}), the tree-level induced FCNCs in the down-type quark processes are only determined by $V^T {\bf y}^q_L$. To reveal the flavor couplings,  ${\bf Y}^{S_1}\equiv V^T {\bf y}^q_L$ can be decomposed as:
 \begin{align}
 Y^{S_1}_d & \approx V_{td} y^q_{L3} -\lambda y^q_{L2} + y^q_{L1}\,, \nonumber \\
 Y^{S_1}_s & \approx V_{ts} y^q_{L3} +  y^q_{L2} + \lambda y^q_{L1}\,, \nonumber \\
 Y^{S_1}_b & \approx  y^q_{L3} \,,
 \end{align}
where we have applied $V_{ud}\approx V_{us}\approx V_{tb} \approx 1$, $V_{us}\approx -V_{cd}\approx\lambda$, and $V_{ts,td}\ll V_{tb}$.  Besides the constraints from the observed loop-mediated processes, such as $K-\bar K$ and $B_q-\bar{B_q}$ mixings, as argued before, the observed $R(D^{(*)})$ excesses also demand that the Yukawa couplings ${\bf y}^q_L$ satisfy the hierarchical structure $|y^q_{L1}| < |y^q_{L2}| < |y^q_{L3}|$. For processes involving the $d_i \to d_j \nu \bar\nu$ transitions, the only relevant new parameters are $m_{S_1}$ and $y^q_{L k}$. We will show that when these parameters are bounded by observables of $\Delta K=2$ and $\Delta B=2$ processes, the model can yield significant deviations on the $B\to K^{(*)} \nu \bar\nu$ and $K\to \pi \nu \bar\nu$ processes from the SM predictions.

Based on the couplings in Eq.~(\ref{eq:LQ_C}), the effective interactions for $d_i \to d_j \nu \bar\nu$, combined with the SM contribution, are given by:
\begin{equation}
{\cal H}_{d_i\to d_j \nu \bar\nu} = C^{\rm SM}_{L} V^*_{t d_i} V_{td_j} \left( X_t + C^{S_1}_{L,ij} \delta_{\ell \tau} \right)  \bar{d_i} \gamma_\mu P_L d_j \, \bar \nu_\ell \gamma^\mu P_L \nu_\ell \,, \label{eq:dsnunu}
\end{equation}
where $P_L=(1-\gamma_5)/2$, $X_t=1.469\pm 0.017$~\cite{Buras:2014fpa}, and the effective coefficients are defined as:
\begin{align}
C^{\rm SM}_L & = \frac{4 G_F}{\sqrt{2}} \frac{\alpha_{\rm em}}{2\pi s^2_W} \,, ~ C^{S_1}_{L,ij}=-\frac{\left( Y^{S_1}_{d_i} \right)^* Y^{S_1}_{d_j}}{2m^2_{S_1} V^*_{t d_i} V_{t d_j} C^{\rm SM}_{L} }\,.
\end{align}
Since only left-handed currents are involved in Eq.~(\ref{eq:dsnunu}), the $S_1$ contributions to the BRs for the decays $B\to M \nu \bar\nu$ with $M=K$ or $K^*$ can be factored out together with the SM result as a multiplicative factor. The resulting BRs can then be simplified as:
\begin{align}
\begin{split}
{\cal B}(B \to M \nu \bar\nu)
&= 
{\cal B}(B \to M \nu \bar\nu)^{\rm SM} R^\nu
\,, \\
\mbox{with }
R^\nu
&= 
\frac{2}{3} + \frac{1}{3} \left| 1 + \frac{C^{S_1}_{L,bs} }{X_t }\right|^2 \,, 
\end{split}
\label{eq:RnuM}
\end{align}
where $B=B^+ (B_d)$ when $M=K^+(K^{*0})$.  Using the $B\to K, K^*$ form factors that combine LCSR and lattice QCD (LQCD)~\cite{Buras:2014fpa,Bharucha:2015bzk,Gubernari:2018wyi} studies,  the SM predictions of their BRs 
are~\cite{Buras:2022wpw}: 
 \begin{align}
 {\cal B}(B^+ \to K^+ \nu \bar\nu)^{\rm SM} = (4.65\pm 0.62) \times 10^{-6}\,,
 \nonumber \\
  {\cal B}(B_d \to K^{*0} \nu \bar\nu)^{\rm SM} =  (10.13\pm 0.92) \times 10^{-6}\,.
 \end{align}
Since the new physics effect can be factored out, the longitudinal polarization fraction of $K^*$ in the model is expected to be the same as that in the SM, i.e.,  $ F^{\rm SM}_L=0.47 \pm 0.03$~\cite{Buras:2014fpa}.  It is interesting to utilize this property to distinguish the interaction structures of potential new physics models.  A formula similar to Eq.~\eqref{eq:RnuM} can be written for the $B \to \pi (\rho) \nu \bar\nu$.  Even though their BRs could have significant deviations from the SM expectations due to the Yukawa coupling $Y_d^{S_1}$, they are still far from current experimental sensitivities.

According to the interactions introduced in Eq.~(\ref{eq:dsnunu}) and the parametrizations for the BRs of the $K^+\to \pi^+ \nu \bar\nu$ and $K_L\to \pi^0 \nu \bar\nu$ decays as shown in Refs.~\cite{Mescia:2007kn,Buras:2015qea}, the influence of $S_1$ on the BRs of these decays can be obtained respectively as:
\begin{align}
\begin{split}
{\cal B}(K^+ \to \pi^+ \nu \bar\nu) 
= &  \frac{2}{3}{\cal B}(K^+ \to \pi^+ \nu \bar\nu)^{\rm SM}  + \frac{\kappa_+ (1+\Delta_{\rm EM}) }{3} \left[ \left( \frac{{\rm Im} X^{S_1}_{\rm eff} }{\lambda^5} \right)^2 \right. \\
& \left. + \left( \frac{{\rm Re}(V^*_{cs} V_{cd} )}{\lambda} P_c (X)+ \frac{{\rm Re}(X^{S_1}_{\rm eff})}{\lambda^5}\right)^2 \right]
\,,  
\end{split}
\label{eq:K_plus}
\end{align}
where  $X^{\rm SM}_{\rm eff}=V^*_{ts} V_{td} X_t$, $X^{S_1}_{\rm eff} = X^{\rm SM}_{\rm eff} +V_{ts}^* V_{td}C^{S_1}_{L,sd}$, $P_c(X)=0.404 \pm 0.024$ denotes the charm-quark contribution~\cite{Isidori:2005xm,Mescia:2007kn,Buras:2015qea}, $\Delta_{\rm EM} = -0.003$, $\kappa_+ =( 5.173 \pm 0.025)\times 10^{-11} (\lambda/0.225)^8$; and
\begin{align}
\begin{split}
{\cal B}(K_L \to \pi^0 \nu \bar\nu) 
= {\cal B}(K_L \to \pi^0\nu \bar\nu)^{\rm SM}  R^\nu
\,.
\end{split}
\label{eq:K_L}
\end{align}
Note that it is a prediction of the model with the assumed hierarchy in the $y^q_{Lk}$ couplings that both ${\cal B}(B \to M \nu \bar\nu)$ and ${\cal B}(K_L \to \pi^0 \nu \bar\nu)$ have approximately the same fractional deviation, $R^\nu$ defined in Eq.~\eqref{eq:RnuM}, from their respective SM values.  The SM predictions for the rare kaon decays are~\cite{Buras:2022wpw}:
\begin{align}
{\cal B}(K^+\to \pi^+ \nu \bar\nu) ^{\rm SM} = (8.60 \pm 0.42)\times 10^{-11}\,, \nonumber \\
%
{\cal B}(K_L\to \pi^0 \nu \bar\nu)^{\rm SM} = (2.94\pm 0.15) \times 10^{-11}\,.
\end{align}
For the $K^+\to \pi^+ \nu \bar\nu$ decay, the current experimental measurement, combining E949 at BNL~\cite{E949:2008btt} and NA62 at CERN~\cite{NA62:2021zjw}, is $(11.4^{+4.0}_{-3.3})\times 10^{-11}$. With the 2021 data analysis by KOTO, the upper limit for $K_L\to \pi^0 \nu \bar\nu$ now is ${\cal B}(K_L\to \pi^0 \nu\bar\nu)< 2 \times 10^{-9}$~\cite{KOTO2023}.  From Eq.~(\ref{eq:K_L}), it can be seen that similar to $B\to M \nu\bar\nu$, the LQ contribution to $K_L\to \pi^0 \nu\bar \nu$ can be expressed as a product of the SM prediction and a scalar factor that encodes the $S_1$ effects.  When the small weak phase of $V_{ts}$ is neglected, $Y^{S_1}_s$ is a real parameter, and the imaginary part of $V_{td}$ from $Y^{S_1}_d$ for $K_L\to \pi^0 \nu \bar\nu$ can be factored out as part of the SM prediction. Consequently, the CP-violating effect does not appear in the multiplicative factor in Eq.~(\ref{eq:K_L}).

The Yukawa coupling $y^q_{L3}$, associated with $y^u_{R2}$, contributes to the $b\to c \tau \nu_\tau$ decay.  To illustrate the influence on $R(D^{(*)})$ in the model, we also show the effective Hamiltonian for $b\to c \ell \bar\nu$ mediated by $W$ and $S^{-1/3}_1$ as~\cite{Chen:2023mep}:
 \begin{align}
 {\cal H}_{b\to c \ell \nu} 
 = &
 \frac{4G_F V_{cb}}{\sqrt{2}} \left[(1 + C^{\ell}_V  \delta^{\ell}_\tau)\bar c \gamma^\mu P_L b~ \bar \ell \gamma_\mu P_L \nu_{\ell} \right. \nonumber \\
 & \left. + C^\tau_S  \bar c P_L b~ \bar\tau P_L \nu_\tau  
 +  C^\tau_T \bar c \sigma_{\mu \nu} P_L  b~ \bar\tau \sigma^{\mu \nu} P_L \nu_\tau \right]
 \,, \label{eq:Hbc}
 \end{align}
where the effective Wilson coefficients at the $m_b$ scale are given by:
  \begin{equation}
  C^{\tau}_V = \frac{\sqrt{2}}{4 G_F V_{cb} }  \frac{y^q_{L3} y^q_{L2}  }{ 2 m^2_{S_1}} \,,~ 
  C^{\tau}_S =  - 1.57 \frac{\sqrt{2} }{4 G_F V_{cb} }  \frac{ y^q_{L3} y^u_{R2}}{2m^2_{S_1}}  
 \,, ~C^{\tau}_T = 0.86\frac{\sqrt{2} }{4G_F V_{cb} } \frac{ y^q_{L3} y^u_{R2} }{8m^2_{S_1}} \,. \label{eq:CRD}
  \end{equation}
It can be seen that although $C^\tau_V$, which has the same current-current interaction structure as the SM, can be induced, yet due to $|y^q_{L2}|  < |y^q_{L3}|$, the dominant effects on $R(D^{(*)})$ by $S_1$ are from the scalar and tensor operators. A detailed study for $R(D)$ and $R(D^*)$ in the model can be found in Ref.~\cite{Chen:2023mep}.

\section{ Constarints from $\Delta K=2$ and $\Delta B=2$} \label{sec:DF=2}

Since the down-type quarks only couple to the left-handed neutrinos via the LQ $S_1$, strict constraints on the parameters $y^q_{L k}$ come from the $\Delta F=2$ processes that are induced via the box diagrams, where $\nu_\tau$ and $S_1$ run in the box loops. Thus, the effective Hamiltonian for $\Delta F=2$ can be derived in a straightforward way as:
\begin{equation}
{\cal H}(\Delta F=2) \approx \frac{1}{2} \frac{(Y^{S^*_1}_{d_i}  Y^{S_1}_{d_j})^2}{(4 \pi)^2 m^2_{S_1}} \left(\bar d_i \gamma_\mu P_L d_j \right)^2\,.
\end{equation}
Using the matrix element $\langle F |( \bar q' \gamma_\mu P_L q)^2 | F \rangle = f^2_F B_F m_F/3$, $\Delta m_K = 2 {\rm Re}(M^K_{12})$ and $\Delta m_{B_q}= 2 |M^B_{12}|$, the mass differneces for $K-\bar K$ and $B_q-\bar B_q$ mixings can be formulated as:
\begin{align}
\Delta m^{S_1}_K = & \frac{{\rm Re}(Y^{S^*_1}_s Y^{S_1}_d )^2}{16\pi^2 m^2_{S_1}} \frac{f^2_K B_K m_K}{3}\,, \nonumber \\
\Delta m^{S_1}_{B_q}= & \frac{|Y^{S^*_1}_b Y^{S_1}_q |^2}{16\pi^2 m^2_{S_1}} \frac{f^2_{B_q} B_{B_q} m_{B_q}}{3}\,, \
\end{align}
where $f_F$ is the decay constant of $F$ meson and $B_F$ is the bag parameter. Since the SM predictions on $\Delta m_K$ and $\Delta m_{B_q}$ are consistent with experimental data and the uncertainties from theoretical non-perturbative QCD effects are larger than those of data, to bound the parameters $y^q_{Lk}$ we conservatively require that the new physics contribution is at least one order of magnitude smaller than the central value of data; that is, we assume 
\begin{equation}
\Delta m^{S_1}_{K,B_q} \lesssim 0.1\, \Delta m^{\rm exp}_{K,B_q}\,, \label{eq:DF=2_data}
\end{equation}
where the current data are $\Delta m^{\rm exp}_{K}=(5.293\pm 0.009)$ (ns)$^{-1}$, $\Delta m^{\rm exp}_{B_d}=0.5065\pm 0.0019$ (ps)$^{-1}$, and $\Delta m^{\rm exp}_{B_s}=(17.765\pm 0.006)$ (ps)$^{-1}$~\cite{ParticleDataGroup:2022pth}.

\begin{figure}[phtb]
\begin{center}
\includegraphics[scale=0.40]{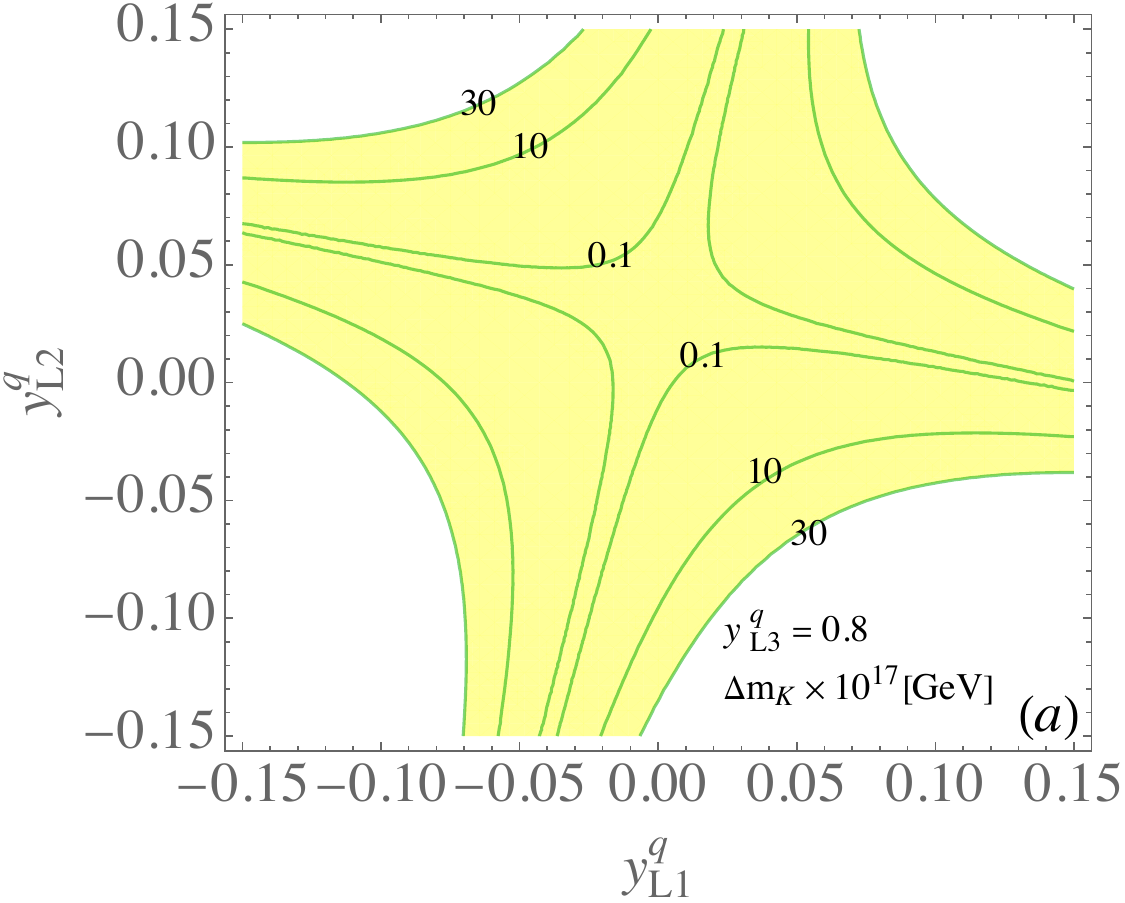}
\hspace{5mm}
\includegraphics[scale=0.40]{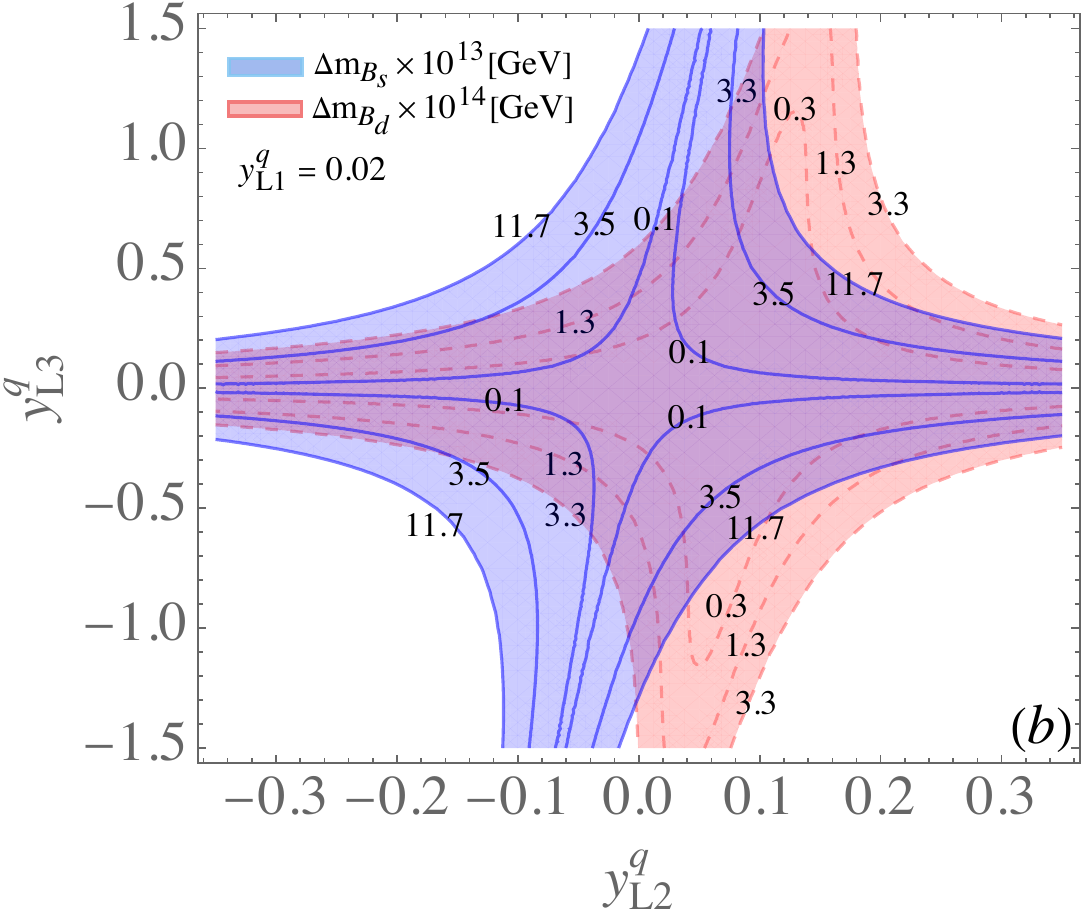}
\caption{(a) $\Delta m^{S_1}_K$ (in units of GeV) with $y^q_{L3}=0.8$ as a function of $y^q_{L1}$ and $y^q_{L2}$. (b) $\Delta m^{S_1}_{B_q}$ (in units of GeV) with $y^q_{L1}=0.02$. In both plots, $\Delta m^{S_1}_{K,B_q} \lesssim 0.1\, \Delta m^{\rm exp}_{K,B_q}$ are applied.}
\label{fig:DF=2}
\end{center}
\end{figure}

To illustrate the $\Delta F=2$ constraints, we show the contours of the mass differences with respect to $y^q_{Lk}$ in Fig.~\ref{fig:DF=2}, where $y^q_{L3}=0.8$ and $y^{q}_{L1}=0.02$ are used for $\Delta m_K$ (left plot) and $\Delta m_{B_q}$ (right plot), respectively. For numerical estimates, we have set $m_{S_1}=1.5$~TeV, $f_{K} \sqrt{B_K}= 0.132$~GeV, $f_{B_d} \sqrt{B_d}= 0.174$~GeV, $f_{B_s} \sqrt{B_s}=0.21$~GeV~\cite{Lenz:2010gu}, $\lambda=0.223$, $V_{td}=A\lambda^2 (1-\rho -i \eta)$ with $A=0.833$, $\rho=0.163$, and $\eta=0.357$, and $V_{ts}=-0.041$~\cite{ParticleDataGroup:2022pth}. From the plots, it is seen that $|y^{q}_{L1}|< |y^{q}_{L2}|< |y^{q}_{L3}|$ is preferred when $|y^{q}_{L3}|\sim O(1)$, as required to explain $R(D^{(*)})$.  We note that the results for $y^q_{L3}=-0.8$ and $y^{q}_{L1}=-0.02$ can be obtained from the corresponding plots in Fig.~\ref{fig:DF=2} by making a parity transformation on the parameters, i.e., $y^q_{L1,L2} \to - y^q_{L1,L2}$ for $\Delta m_K$ and $y^q_{L2,L3} \to - y^q_{L2,L3}$ for $\Delta m_{B_q}$.  We also note in passing that the parameter distributions for $\Delta m_K$ and $\Delta m_{B_d}$ do not center at the origin in the respective plots because of our choices of $y^q_{L3}=0.8$ and $y^q_{L1}=0.02$.  The distribution for $\Delta m_{B_s}$ in plot (b), on the other hand, is symmetric with respect to the origin because it is insensitive to the choice of $y^q_{L1}$.

\section{Numerical results and discussions}\label{sec:NA}

In this section, we analyze the $S_1$ contributions to the $B\to M \nu \bar\nu$ and $K\to \pi \nu \bar\nu$ decays when the constraints from $\Delta F=2$ are all taken into account. In this model, the LQ only couples to the third-generation lepton, and the involved parameters in the model are $y^q_{Lk}$ and $m_{S_1}$.   Both CMS~\cite{CMS:2020wzx} and ATLAS~\cite{ATLAS:2021oiz} have searched for the scalar LQ with a charge of $e/3$ using the $t\tau$ and $b\nu$ production channels.  An upper bound on the LQ mass is given by ATLAS to be $m_{S} \ge 1.22$~TeV when $BR(S_1 \to t \tau)=1/2$.  Given this measurement and the allowed parameter ranges in Fig.~\ref{fig:DF=2}, $t\tau$ and $b\nu_\tau$ are the dominant decays of the LQ in the model. Thus, we take $m_{S_1}=1.5$~TeV in our numerical calculations.  Since $|y^q_{L3}|$  as large as $1.5$ for $m_{S_1}=1.5$~TeV is still not excluded by the current data, we therefore consider $|y^q_{L3}|\lesssim 1.5$ in the following analysis.  Using the high-$p_T$ tail of the $pp\to \tau\tau$ distribution measured by ATLAS~\cite{ATLAS:2020zms} with the integrated luminosity of 139 fb$^{-1}$, the bound on the $c$-$\tau$-$S_1$ coupling can be obtained as $|y^u_{R2}|\lesssim 1.6$~\cite{Angelescu:2021lln}.

To determine the parameter space of the three parameters $y^q_{Lk}$ under the constraints of $\Delta F=2$ processes, we perform a random parameter scan within the following ranges:
\begin{equation}
y^{q}_{L1} \in (-0.05, 0.05)\,,~ y^q_{L2} \in (-0.15, 0.15)\,,~ y^q_{L3} \in (-1.5,1.5)\,.
\label{eq:y-ranges}
\end{equation}
The ranges of $\Delta m^{S_1}_{K,B_q}$ that satisfy the conditions in Eq.~(\ref{eq:DF=2_data}) are explicitly taken as follows: $\Delta m^{S_1}_{K} \in (0.01, 34.8)\times 10^{-17}$ GeV,  $\Delta m^{S_1}_{B_d} \in (0.01, 33.3)\times 10^{-15}$ GeV, and $ \Delta m^{S_1}_{B_s} \in (0.01, 11.7)\times 10^{-13}$ GeV.  
Besides, we will require that the predicted ${\cal B}(K^+\to \pi^+ \nu \bar\nu)$ fall within its $\pm 1\sigma$ range.

\begin{figure}[phtb]
\begin{center}
\includegraphics[scale=0.55]{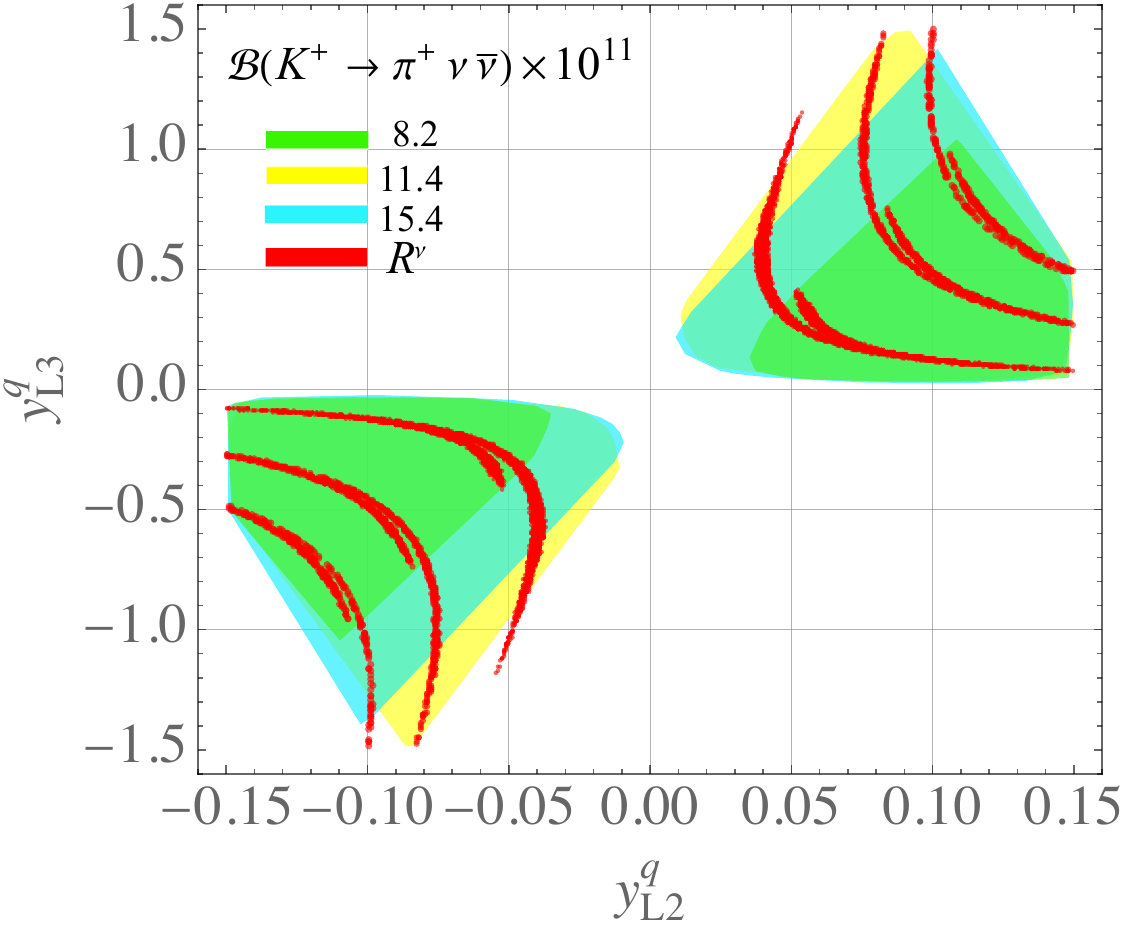}
\caption{Predictions for ${\cal B}(K^+ \to \pi^+ \nu \bar\nu)$ and $R^\nu$ in the plane of $y^q_{L2}$ and $y^q_{L3}$, with $y^q_{L1}$ varying within the specified range and under the constraints discussed in the main text.  }
\label{fig:B_K}
\end{center}
\end{figure}

Using $10^{7}$ sampling points and the constraints mentioned above, the predicted BR for $K^+\to \pi^+ \nu \bar\nu$ in the $y^q_{L2}$-$y^q_{ L3}$ plane is shown in Fig.~\ref{fig:B_K}, where the green, yellow, and cyan regions give the BRs of $(8.2, 11.4, 15.4) \times 10^{-11}$, respectively.  The reason for such a spreading pattern for each specific BR is because of the more intricate dependence of $y^q_{Lk}$ in ${\cal B}(K^+\to \pi^+\nu\bar\nu)$, as revealed in Eq.~\eqref{eq:K_L}, than that in ${\cal B}(B\to M \nu \bar\nu)$ and ${\cal B}(K_L\to \pi^0 \nu \bar\nu)$.  It is also because of this observable that, compared to considering only $\Delta m_{K, B_d, B_s}$ as the examples in Fig.~\ref{fig:DF=2}, the preferred parameter space in the plane is restricted to the first and third quadrants.  Note that the parameter space around the origin is excluded because we have assumed minimum new physics contributions to $\Delta m^{S_1}_{K,B_d, B_s}$.  This is also required in order to have significant deviations in ${\cal B}(B\to M \nu \bar\nu)$ and ${\cal B}(K_L\to \pi^0 \nu \bar\nu)$.

As alluded to before, the $S_1$ contributions to ${\cal B}(B\to M \nu \bar\nu)$ and ${\cal B}(K_L\to \pi^0 \nu \bar\nu)$ can be factored out together with the SM contributions into a scalar factor characterized by $R^\nu$ defined in Eq.~\eqref{eq:RnuM}.  We superimpose the distribution for $R^\nu=(1.2,1.8,2.6)$ in Fig.~\ref{fig:B_K} to show that such values are consistent with the current measurement of ${\cal B}(K^+\to \pi^+\nu\bar\nu)$.  The dispersion in each particular value of $R^\nu$ is due to the variation in $y^q_{L1}$.  This means that the BRs of ${\cal B}(B\to M \nu \bar\nu)$ and ${\cal B}(K_L\to \pi^0 \nu \bar\nu)$ are allowed to be enhanced by a factor of 2 or more, thus accommodating the Belle~II data in Eq.~\eqref{eq:BelleII-data}.

\begin{figure}[phtb]
\begin{center}
\includegraphics[scale=0.35]{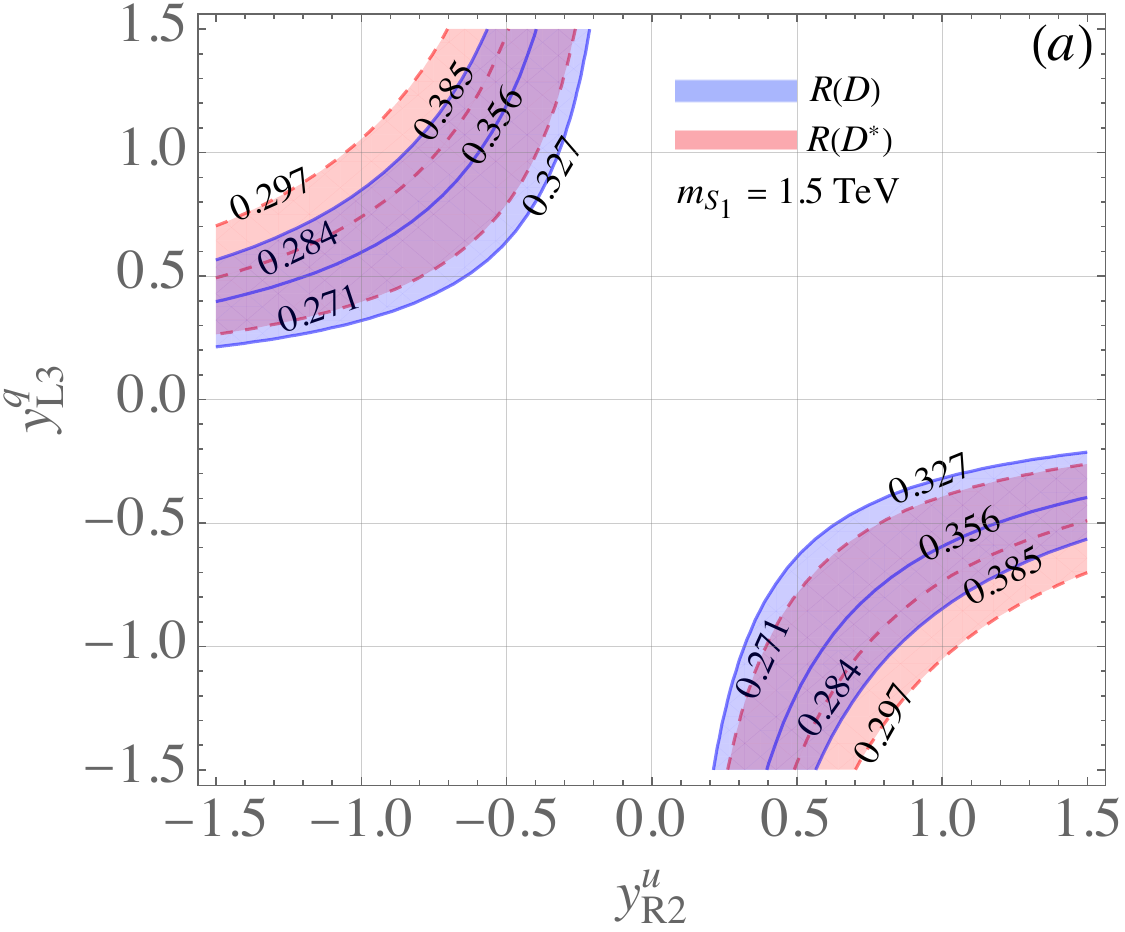} 
\hspace{5mm}
\includegraphics[scale=0.35]{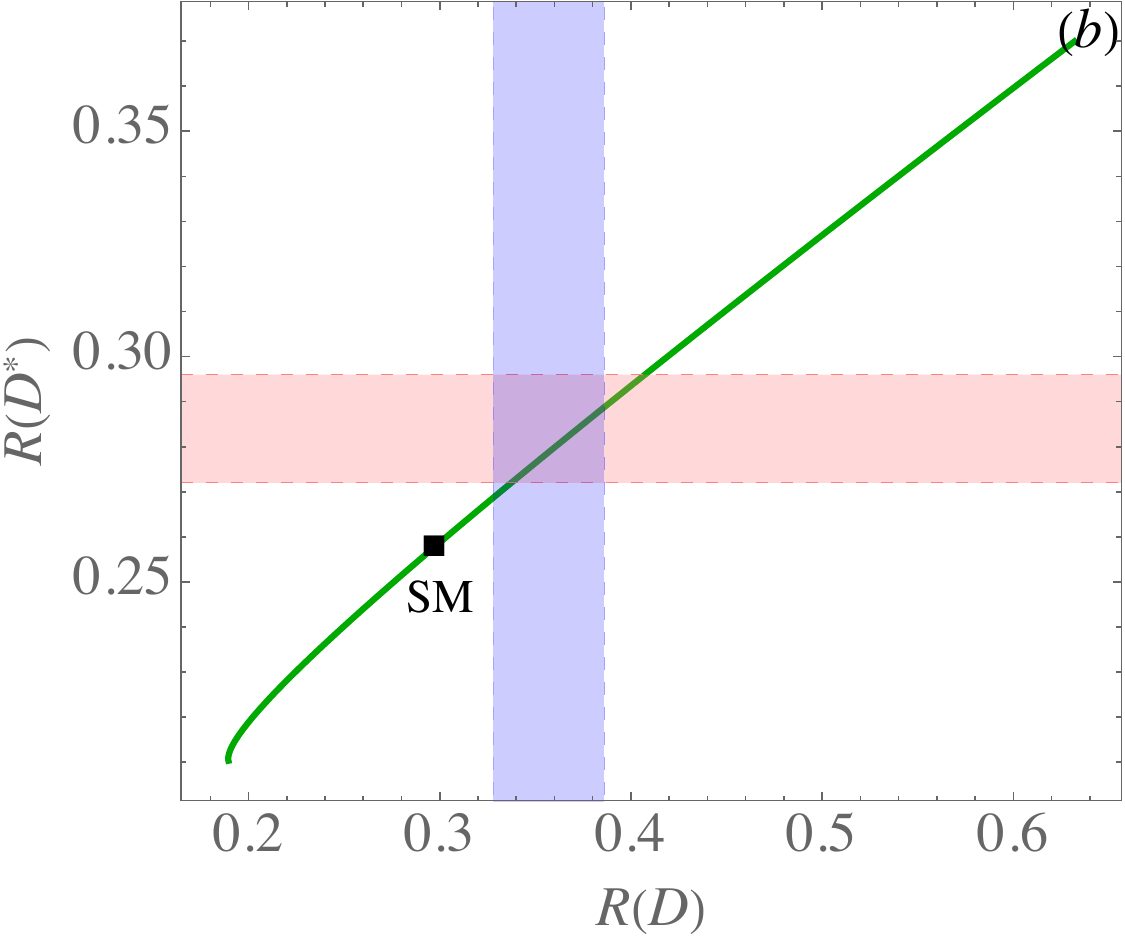} 
\caption{ (a) Contours of $R(D^{(*)})$ in the $y^u_{R2}$-$y^q_{L3}$ plane, and (b) correlation between $R(D)$ and $R(D^*)$ in the model.  The light blue and pink bands represent the $\pm 1\sigma$ bands of $R(D)$ and $R(D^*)$, respectively.  The solid square marks the SM predictions.  }
\label{fig:RD}
\end{center}
\end{figure}

Finally, we comment on the impact of $S_1$ on the observables $R(D)$ and $R(D^*)$ in the model.  Neglecting the minor influence of $C^\tau_V$ in Eq.~(\ref{eq:CRD}), the parameters involved in the $b\to c \tau \nu$ transition appear in the combination $y^q_{L3} y^u_{R2}/m^2_{S_1}$. 
 Using the formulae given in Ref.~\cite{Chen:2023mep}, we show in Fig.~\ref{fig:RD}(a) several contours of $R(D)$ and $R(D^*)$ in the plane of $y^u_{R2}$ and $y^q_{L3}$ for the case of $m_{S_1}=1.5$~TeV.  The $\pm 1\sigma$ ranges of $R(D)$ and $R(D^*)$ data are seen to have a significant overlap.  The correlation between $R(D)$ and $R(D^*)$ as we vary the value of the dominant factor $y^q_{L3} y^u_{R2}$ is shown in Fig.~\ref{fig:RD}(b), where the SM predictions, $R^{\mathrm{SM}}(D) \approx 0.297$ and $R^{\mathrm{SM}}\left(D^*\right) \approx 0.258$~\cite{Chen:2023mep}, are marked by the black square. Due to the absence of a significant interfering effect between the SM and $S_1$ contributions, the linear relationship between $R(D)$ and $R(D^*)$ does not depend on the values of the parameters involved (e.g., $m_{S_1}$).  The fact that the predicted correlation curve goes through a good portion of the crossed region reflects the overlapped parameter space in Fig.~\ref{fig:RD}(a).  A more precise determination of both $R(D)$ and $R(D^*)$ will be able to show whether they are still in line with the model predictions.

\section{Summary \label{sec:summary}}

We impose the $U(1)_{\mu-\tau}$ gauge symmetry to the $S_1$ LQ model, where the Yukawa couplings to the LQ are greatly simplified.  In addition to the muon $g-2$ anomaly, we have found that the predicted $R(D)$ and $R(D^*)$ values can explain the excesses indicated by the experimental data. Moreover, given a fixed LQ mass, the $d_i \to d_j \nu \bar\nu$ processes depend only on the three Yukawa couplings $y^q_{Lk}$. When considering constraints from $\Delta m_K$ and $\Delta m_{B_d,B_s}$, the processes $B\to K (K^*) \nu \bar\nu$, $K^+\to \pi ^+\nu \bar\nu$, and $K_L\to \pi^0 \nu \bar\nu$ can be significantly enhanced by the effects of $S_1$.  Such enhancements can be readily tested in future experiments.  We emphasize that the KM phase is assumed to be the sole source of CP violation in the study; nevertheless, the CP-violating process $K_L\to \pi^0 \nu \bar\nu$ can still be enhanced by at least a factor of 2 compared to the SM prediction.

\acknowledgments
This work was supported in part by the National Science and Technology Council, Taiwan under Grant Nos. MOST-110-2112-M-006-010-MY2 (C.-H.~Chen) and MOST-111-2112-M-002-018-MY3 (C.-W.~Chiang).



\begin{thebibliography}{99}


\bibitem{Muong-2:2023cdq}
D.~P.~Aguillard \textit{et al.} [Muon g-2],
[arXiv:2308.06230 [hep-ex]].


\bibitem{Aoyama:2020ynm}
T.~Aoyama, N.~Asmussen, M.~Benayoun, J.~Bijnens, T.~Blum, M.~Bruno, I.~Caprini, C.~M.~Carloni Calame, M.~C\`e and G.~Colangelo, \textit{et al.}
Phys. Rept. \textbf{887}, 1-166 (2020)
[arXiv:2006.04822 [hep-ph]].


\bibitem{EPS_Belle2a} Eldar Ganiev, on behalf of Belle II Collaboration, talk 
presented at EPS-HEP, 21-25 August 2023, Germany. 

\bibitem{BaBar:2013npw}
J.~P.~Lees \textit{et al.} [BaBar],
Phys. Rev. D \textbf{87}, no.11, 112005 (2013)
[arXiv:1303.7465 [hep-ex]].


\bibitem{Belle:2013tnz}
O.~Lutz \textit{et al.} [Belle],
Phys. Rev. D \textbf{87}, no.11, 111103 (2013)
[arXiv:1303.3719 [hep-ex]].


\bibitem{Belle:2017oht}
J.~Grygier \textit{et al.} [Belle],
Phys. Rev. D \textbf{96}, no.9, 091101 (2017)
[arXiv:1702.03224 [hep-ex]].

\bibitem{Belle-II:2021rof}
F.~Abudin\'en \textit{et al.} [Belle-II],
Phys. Rev. Lett. \textbf{127}, no.18, 181802 (2021)
[arXiv:2104.12624 [hep-ex]].



\bibitem{Buras:2022wpw}
A.~J.~Buras and E.~Venturini,
Eur. Phys. J. C \textbf{82}, no.7, 615 (2022)
[arXiv:2203.11960 [hep-ph]].


\bibitem{Buras:2014fpa}
A.~J.~Buras, J.~Girrbach-Noe, C.~Niehoff and D.~M.~Straub,
JHEP \textbf{02}, 184 (2015)
[arXiv:1409.4557 [hep-ph]].



\bibitem{Browder:2021hbl}
T.~E.~Browder, N.~G.~Deshpande, R.~Mandal and R.~Sinha,
Phys. Rev. D \textbf{104}, no.5, 053007 (2021)
[arXiv:2107.01080 [hep-ph]].

\bibitem{Asadi:2023ucx}
P.~Asadi, A.~Bhattacharya, K.~Fraser, S.~Homiller and A.~Parikh,
[arXiv:2308.01340 [hep-ph]].


\bibitem{Athron:2023hmz}
P.~Athron, R.~Martinez and C.~Sierra,
[arXiv:2308.13426 [hep-ph]].


\bibitem{Bause:2023mfe}
R.~Bause, H.~Gisbert and G.~Hiller,
[arXiv:2309.00075 [hep-ph]].

\bibitem{Allwicher:2023syp}
L.~Allwicher, D.~Becirevic, G.~Piazza, S.~Rosauro-Alcaraz and O.~Sumensari,
[arXiv:2309.02246 [hep-ph]].

\bibitem{Felkl:2023ayn}
T.~Felkl, A.~Giri, R.~Mohanta and M.~A.~Schmidt,
[arXiv:2309.02940 [hep-ph]].

\bibitem{Dreiner:2023cms}
H.~K.~Dreiner, J.~Y.~G\"unther and Z.~S.~Wang,
[arXiv:2309.03727 [hep-ph]].

\bibitem{Amhis:2023mpj}
Y.~Amhis, M.~Kenzie, M.~Reboud and A.~R.~Wiederhold,
[arXiv:2309.11353 [hep-ex]].


\bibitem{Fajfer:2012jt}
S.~Fajfer, J.~F.~Kamenik, I.~Nisandzic and J.~Zupan,
Phys. Rev. Lett. \textbf{109}, 161801 (2012)
[arXiv:1206.1872 [hep-ph]].

\bibitem{Sakaki:2013bfa} 
  Y.~Sakaki, M.~Tanaka, A.~Tayduganov and R.~Watanabe,
  Phys.\ Rev.\ D {\bf 88}, no. 9, 094012 (2013)
  [arXiv:1309.0301 [hep-ph]].


\bibitem{Calibbi:2015kma}
L.~Calibbi, A.~Crivellin and T.~Ota,
Phys. Rev. Lett. \textbf{115}, 181801 (2015)
[arXiv:1506.02661 [hep-ph]].

\bibitem{Sahoo:2015qha} 
  S.~Sahoo and R.~Mohanta,
  Phys.\ Rev.\ D {\bf 93}, no. 3, 034018 (2016)
  [arXiv:1507.02070 [hep-ph]].
  
\bibitem{Chen:2017hir}
C.~H.~Chen, T.~Nomura and H.~Okada,
Phys. Lett. B \textbf{774}, 456-464 (2017)
[arXiv:1703.03251 [hep-ph]].


\bibitem{Crivellin:2019dwb}
A.~Crivellin, D.~M\"uller and F.~Saturnino,
JHEP \textbf{06}, 020 (2020)
[arXiv:1912.04224 [hep-ph]].


\bibitem{Davighi:2020qqa}
J.~Davighi, M.~Kirk and M.~Nardecchia,
JHEP \textbf{12}, 111 (2020)
[arXiv:2007.15016 [hep-ph]].


\bibitem{Greljo:2021xmg}
A.~Greljo, P.~Stangl and A.~E.~Thomsen,
Phys. Lett. B \textbf{820}, 136554 (2021)
[arXiv:2103.13991 [hep-ph]].


\bibitem{Carvunis:2021dss}
A.~Carvunis, A.~Crivellin, D.~Guadagnoli and S.~Gangal,
Phys. Rev. D \textbf{105}, no.3, L031701 (2022)
[arXiv:2106.09610 [hep-ph]].


\bibitem{Davighi:2022qgb}
J.~Davighi, A.~Greljo and A.~E.~Thomsen,
Phys. Lett. B \textbf{833}, 137310 (2022)
[arXiv:2202.05275 [hep-ph]].



\bibitem{Heeck:2022znj}
J.~Heeck and A.~Thapa,
Eur. Phys. J. C \textbf{82}, no.5, 480 (2022)
[arXiv:2202.08854 [hep-ph]].





\bibitem{ParticleDataGroup:2022pth}
R.~L.~Workman \textit{et al.} [Particle Data Group],
PTEP \textbf{2022}, 083C01 (2022).

\bibitem{HeavyFlavorAveragingGroup:2022wzx}
Y.~S.~Amhis \textit{et al.} [Heavy Flavor Averaging Group and HFLAV],
Phys. Rev. D \textbf{107}, no.5, 052008 (2023)
[arXiv:2206.07501 [hep-ex]].

\bibitem{LHCb:2022vje}
R. Aaij {\it et. al} [LHCb],
Phys. Rev. D \textbf{108}, no.3, 032002 (2023)
[arXiv:2212.09153 [hep-ex]].


\bibitem{Buras:2022qip}
A.~J.~Buras,
Eur. Phys. J. C \textbf{83}, no.1, 66 (2023)
[arXiv:2209.03968 [hep-ph]].



\bibitem{MILC:2015uhg}
J.~A.~Bailey \textit{et al.} [MILC],
Phys. Rev. D \textbf{92}, no.3, 034506 (2015)
[arXiv:1503.07237 [hep-lat]].

\bibitem{Na:2015kha}
H.~Na \textit{et al.} [HPQCD],
Phys. Rev. D \textbf{92}, no.5, 054510 (2015)
[erratum: Phys. Rev. D \textbf{93}, no.11, 119906 (2016)]
[arXiv:1505.03925 [hep-lat]].


  
\bibitem{Bigi:2016mdz}
D.~Bigi and P.~Gambino,
Phys. Rev. D \textbf{94}, no.9, 094008 (2016)
[arXiv:1606.08030 [hep-ph]].

\bibitem{Bernlochner:2017jka}
F.~U.~Bernlochner, Z.~Ligeti, M.~Papucci and D.~J.~Robinson,
Phys. Rev. D \textbf{95}, no.11, 115008 (2017)
[erratum: Phys. Rev. D \textbf{97}, no.5, 059902 (2018)]
[arXiv:1703.05330 [hep-ph]].



\bibitem{Jaiswal:2017rve}
S.~Jaiswal, S.~Nandi and S.~K.~Patra,
JHEP \textbf{12}, 060 (2017)
[arXiv:1707.09977 [hep-ph]].

\bibitem{BaBar:2019vpl}
J.~P.~Lees \textit{et al.} [BaBar],
Phys. Rev. Lett. \textbf{123}, no.9, 091801 (2019)
[arXiv:1903.10002 [hep-ex]].

\bibitem{Bordone:2019vic}
M.~Bordone, M.~Jung and D.~van Dyk,
Eur. Phys. J. C \textbf{80}, no.2, 74 (2020)
[arXiv:1908.09398 [hep-ph]].

\bibitem{Martinelli:2021onb}
G.~Martinelli, S.~Simula and L.~Vittorio,
Phys. Rev. D \textbf{105}, no.3, 034503 (2022)
[arXiv:2105.08674 [hep-ph]].



\bibitem{He:1991qd}
X.~G.~He, G.~C.~Joshi, H.~Lew and R.~R.~Volkas,
Phys. Rev. D \textbf{44}, 2118-2132 (1991).

\bibitem{Heeck:2011wj}
J.~Heeck and W.~Rodejohann,
Phys. Rev. D \textbf{84}, 075007 (2011)
[arXiv:1107.5238 [hep-ph]].

\bibitem{Chen:2017cic}
C.~H.~Chen and T.~Nomura,
Phys. Rev. D \textbf{96}, no.9, 095023 (2017)
[arXiv:1704.04407 [hep-ph]].


\bibitem{Altmannshofer:2014cfa}
W.~Altmannshofer, S.~Gori, M.~Pospelov and I.~Yavin,
Phys. Rev. D \textbf{89}, 095033 (2014)
[arXiv:1403.1269 [hep-ph]].


\bibitem{Altmannshofer:2014pba}
W.~Altmannshofer, S.~Gori, M.~Pospelov and I.~Yavin,
Phys. Rev. Lett. \textbf{113}, 091801 (2014)
[arXiv:1406.2332 [hep-ph]].


\bibitem{Altmannshofer:2016oaq}
W.~Altmannshofer, M.~Carena and A.~Crivellin,
Phys. Rev. D \textbf{94}, no.9, 095026 (2016)
[arXiv:1604.08221 [hep-ph]].



\bibitem{Chen:2023mep}
C.~H.~Chen, C.~W.~Chiang and C.~W.~Su,
[arXiv:2305.09256 [hep-ph]].





\bibitem{Bharucha:2015bzk}
A.~Bharucha, D.~M.~Straub and R.~Zwicky,
JHEP \textbf{08}, 098 (2016)
[arXiv:1503.05534 [hep-ph]].

\bibitem{Gubernari:2018wyi}
N.~Gubernari, A.~Kokulu and D.~van Dyk,
JHEP \textbf{01}, 150 (2019)
[arXiv:1811.00983 [hep-ph]].


\bibitem{Mescia:2007kn}
F.~Mescia and C.~Smith,
Phys. Rev. D \textbf{76}, 034017 (2007)
[arXiv:0705.2025 [hep-ph]].

\bibitem{Buras:2015qea}
A.~J.~Buras, D.~Buttazzo, J.~Girrbach-Noe and R.~Knegjens,
JHEP \textbf{11}, 033 (2015)
[arXiv:1503.02693 [hep-ph]].

\bibitem{Isidori:2005xm} 
  G.~Isidori, F.~Mescia and C.~Smith,
  Nucl.\ Phys.\ B {\bf 718}, 319 (2005)
  [hep-ph/0503107].
  



\bibitem{Lenz:2010gu}
A.~Lenz, U.~Nierste, J.~Charles, S.~Descotes-Genon, A.~Jantsch, C.~Kaufhold, H.~Lacker, S.~Monteil, V.~Niess and S.~T'Jampens,
Phys. Rev. D \textbf{83}, 036004 (2011)
[arXiv:1008.1593 [hep-ph]].


\bibitem{CMS:2020wzx}
A.~M.~Sirunyan \textit{et al.} [CMS],
Phys. Lett. B \textbf{819}, 136446 (2021)
[arXiv:2012.04178 [hep-ex]].

\bibitem{ATLAS:2021oiz}
G.~Aad \textit{et al.} [ATLAS],
JHEP \textbf{06}, 179 (2021)
[arXiv:2101.11582 [hep-ex]].


\bibitem{ATLAS:2020zms}
G.~Aad \textit{et al.} [ATLAS],
Phys. Rev. Lett. \textbf{125}, no.5, 051801 (2020)
[arXiv:2002.12223 [hep-ex]].

\bibitem{Angelescu:2021lln}
A.~Angelescu, D.~Be\v{c}irevi\'c, D.~A.~Faroughy, F.~Jaffredo and O.~Sumensari,
Phys. Rev. D \textbf{104}, no.5, 055017 (2021)
[arXiv:2103.12504 [hep-ph]].


\bibitem{E949:2008btt}
A.~V.~Artamonov \textit{et al.} [E949],
Phys. Rev. Lett. \textbf{101}, 191802 (2008)
[arXiv:0808.2459 [hep-ex]].

\bibitem{NA62:2021zjw}
E.~Cortina Gil \textit{et al.} [NA62],
JHEP \textbf{06}, 093 (2021)
[arXiv:2103.15389 [hep-ex]].

\bibitem{KOTO2023} 
Koji Shiomi, on behalf of  KOTO Collaboration, talk 
presented at KEK IPNS and J-PARC Joint Seminar, 6th Sep 2023. 


\end{thebibliography}
\end{document}